%% file: paper.tex
\documentclass{aa}
\usepackage{graphicx,natbib}
\usepackage{txfonts}
\usepackage{epic,eepic}

\begin{document}
 
\input macros
\input aas_journals

\baselineskip=12pt

\title{A 3D radiative transfer framework: II. line transfer problems}

\titlerunning{3D radiative transfer framework II}
\authorrunning{Baron and Hauschildt}
\author{E.~Baron\inst{1,2,3} and Peter H. Hauschildt\inst{1}}

\institute{
Hamburger Sternwarte, Gojenbergsweg 112, 21029 Hamburg, Germany;
yeti@hs.uni-hamburg.de 
\and
Dept. of Physics and Astronomy, University of
Oklahoma, 440 W.  Brooks, Rm 100, Norman, OK 73019 USA;
baron@nhn.ou.edu
\and
NERSC, Lawrence Berkeley National Laboratory, MS
50F-1650, 1 Cyclotron Rd, Berkeley, CA 94720-8139 USA
}

\date{Received date \ Accepted date}

\abstract
{Higher resolution telescopes as well as 3D numerical simulations
  will require the development of detailed 3D radiative transfer
  calculations. Building upon our previous work we extend our method
  to include both continuum and line transfer.}
{We present a general method to calculate radiative transfer including
scattering in the continuum as well as in lines in 3D static atmospheres.
}
{The scattering problem for line transfer is solved via means of an
  operator splitting (OS) technique. The formal solution
  is based on a long-characteristics method. The approximate
  $\Lambda$ operator is constructed considering nearest
  neighbors {\em exactly}. The code is parallelized over both wavelength and solid angle
  using the MPI library.
}
{We present the results of several test cases with different values of
  the thermalization parameter and two choices for the temperature
  structure. The results are directly compared to 1D spherical
  tests. With our current grid setup the interior resolution is much
  lower in 3D than in 1D, nevertheless the 3D results agree very
  well with the well-tested 1D calculations. We show that with
  relatively simple parallelization that the code scales to very large
  number of processors which is mandatory  for practical applications.
}
{Advances in modern computers will make realistic 3D radiative transfer
  calculations possible in the near future. Our current code scales to
very large numbers of processors, but requires larger memory per
processor at high spatial resolution.}


\maketitle

\section{Introduction}

Hydrodynamical calculations in two and three spatial dimensions are
necessary in a broad range of astrophysical contexts. With modern
parallel supercomputers, they are also becoming more realistic, in
that they can be run at modest to high resolution.
Performing full radiation hydrodynamical calculations is presently still too
computationally expensive.
Recently, \citet{HB06} have presented a mixed
frame method of solving the time-dependent radiative transfer problem
in 2D, but their work is tailored toward neutrino transport where the
absence of rapidly changing opacity such as a spectral line makes their
approximations appropriate. Similar work has been presented by
\citet{MK82}, \citet{lowrie99}, and \citet{lowrie01}. Taking a
different approach \citet{KKM06} derive the flux-limiter to $O(v/c)$
for use in radiation hydrodynamics calculations. Similar work was
presented by \citet{cb92}. 
Even though these recent first steps are improvements, they suffer
from a loss of accuracy, either in dealing with spectral lines, or in
obtaining the correct angular dependence of the photon distribution
function or the specific intensity. 
While these recent works are expedient, they are crude enough that the
results of the hydrodynamical calculations cannot be compared directly
with observed spectra. Given the fact that computational resources are
finite, a final post-processing step is necessary to compare the
results of hydrodynamical calculations to observations. 

In \citet[][hereafter: Paper I]{3drt_paper1}  we described a framework for
the solution of the radiative transfer equation for scattering continua
in 3D (when we say 3D we mean three spatial dimensions, plus three
momentum dimensions) for 
the time independent, static case. Here we extend our method to
include transfer in lines including the case that the line 
is scattering dominated. {\citet{BTB97} presented a
multi-level, multi-grid, multi-dimensional radiative transfer scheme,
using a lower triangular ALO and solving the scattering problem via a
Gauss-Seidel method.  \citet*{vannoort02} presented a method of
solving the full NLTE radiative transfer problem using the short
characteristics method in 2-D for Cartesian, spherical, and
cylindrical geometry. They also used the technique of accelerated
lambda iteration (ALI) \citep{OAB,ok87}, however they restricted
themselves to the case of a diagonal accelerated lambda operator
(ALO).}

{We describe our method, its rate of convergence, and present
  comparisons to our 
  well-tested 1-D calculations.}

\section{Method}

In the following discussion we use  notation of \citet{s3pap} and Paper I.  The
basic framework and the methods used for the formal solution and the solution
of the scattering problem via operator splitting are discussed in detail in
paper I and will thus not be repeated here. We have extended the framework 
to solve line transfer problems with a background continuum. The basic 
approach is similar to that of \cite{casspap}. In the simple case of a 
2-level atom with background continuum we consider here as a test case, 
we use a wavelength grid that covers the profile of the line including the 
surrounding continuum. We then use the wavelength dependent mean intensities
$J_\lambda$ and approximate $\Lambda$ operators $\Lstar$ to compute the profile integrated
line mean intensities $\Jb$ and $\bar\Lstar$ via
\[
\Jb = \int \phi(\lambda) J_\lambda\,d\lambda
\]
and 
\[
\bar\Lstar = \int \phi(\lambda) \Lstar\,d\lambda.
\]
$\Jb$ and $\bar\Lstar$ are then used to compute an updated value for 
$\Jb$ and the line source function 
\[ S = (1-\epsilon)\Jb+\epsilon B \]
where $\epsilon$ is the line thermalization parameter ($0$ for a purely
absorptive line, $1$ for a purely scattering line). $B$ is the Planck 
function, $B_\lambda$, profile averaged over the line 
\[ B = \int \phi(\lambda) B_\lambda\,d\lambda \]
via the standard iteration method 
\[
    \left[1-\lstar(1-\e)\right]\bar\Jnew = \bar\Jfs -\lstar(1-\e)\bar\Jold, 
\]
where $\Jfs=\bar\L\Sold$. This equation is solved directly to get the new values of 
$\Jb$ which is then used to compute the new 
source function for the next iteration cycle.

We construct the line $\bar\Lstar$ directly from the wavelength
dependent $\Lstar$'s generated by the solution of the continuum
transfer problems.  For practical reasons, we use in this paper only
the nearest neighbor $\Lstar$ discussed in paper I. Larger $\Lstar$s
require significantly more storage and small test cases indicate that
they do not decrease the number of iterations enough to warrant their
use as long as they are not much larger than the nearest
neighbor $\Lstar$.

\section{Application examples}

We use the framework discussed in paper I as the baseline for the
line transfer problems discussed in this paper. In addition to 
the highly efficient parallelization of solid angle space, we 
have implemented a parallelization over wavelength space using 
the MPI distributed memory model. For static configurations (or 
for configurations with velocity fields treated in the Eulerian frame) 
there is no direct coupling between different wavelength points.

Our basic setup is similar to that discussed in paper I.  We use a sphere with
a grey continuum opacity parameterized by a power law in the continuum optical
depth $\tstd$. The basic model parameters are
\begin{enumerate}
\item Inner radius $r_c=10^{13}\,$cm, outer radius $\rout = 1.01\alog{15}\,$cm.
\item Minimum optical depth in the continuum $\t_{\rm std}^{\rm min} =
10^{-4}$ and maximum optical depth in the continuum $\t_{\rm std}^{\rm
max} = 1$.
\item constant temperature structure with $T=10^4$~K {\em or}
\item grey temperature structure with $\Teff=10^4$~K.
\item Outer boundary condition $I_{\rm bc}^{-} \equiv 0$ and diffusion
inner boundary condition for all wavelengths.
\item Continuum extinction $\chi_c = C/r^2$, with the constant $C$
fixed by the radius and optical depth grids.
\item Parameterized coherent \& isotropic continuum scattering by
defining
\[
\chi_c = \epsilon_c \kappa_c + (1-\epsilon_c) \sigma_c
\]
with $0\le \epsilon_c \le 1$. 
$\kappa_c$ and $\sigma_c$ are the
continuum absorption and scattering coefficients.
\end{enumerate}

The line of the simple 2-level model atom is parameterized by the ratio of the
profile averaged  line opacity $\chi_l$ to the continuum opacity $\chi_c$ and
the line thermalization parameter $\epsilon_l$. For the test cases presented
below, we have used $\epsilon_c=1$ and a constant temperature and thus a
constant thermal part of the source function for simplicity (and to save
computing time) and set $\chi_l/\chi_c = 10^6$ to simulate a strong line, with
varying $\epsilon_l$ (see below). With this setup, the optical depths as seen
in the line range from $10^{-2}$ to $10^6$. We use 32 wavelength points to model
the full line profile, including wavelengths outside the line for the
continuum.  We did not require the line to thermalize at the center of the test
configurations, this is a typical situation one encounters in a full 3D
configurations as the location (or even existence) of the
thermalization depths becomes more 
ambiguous than in the 1D case.

The sphere is put at the center of the  Cartesian grid, which is 
in each axis 10\% larger than the radius of the sphere.
For the test calculations we use voxel grids with the same
number of spatial points in each direction (see below). The
solid angle space was discretized in $(\theta,\phi)$ with 
$n_\theta=n_\phi$ if not stated otherwise. In the following 
we discuss the results of various tests. In all tests we use
the LC method for the 3D RT solution. Unless otherwise stated,
the tests were run on parallel computers using 128 CPUs. For the 3D solver
we use $n_x=n_y=n_z=2*64+1$ or $n_x=n_y=n_z=2*96+1$ points along each axis, for
a total of $129^3$ or  $193^3$ spatial points, depending on the test case. The
solid angle space discretization uses $n_\theta=n_\phi=64$ points.

\subsection{LTE tests}
\begin{figure}
\centering
\resizebox{\hsize}{!}{\includegraphics[angle=90]{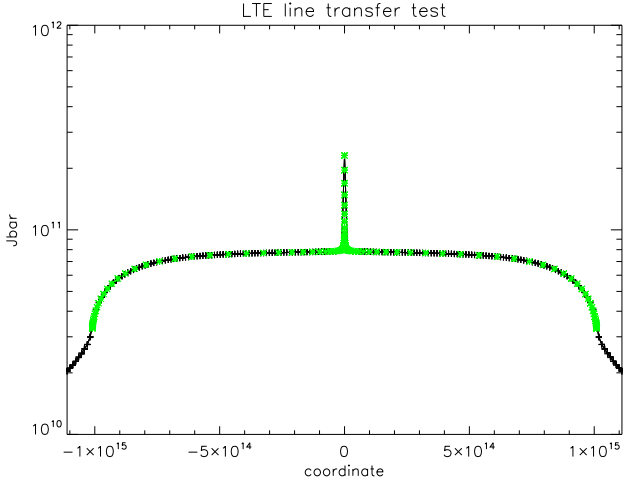}}
\caption{\label{fig:LTE:axes} Comparison of the results obtained 
for the LTE line  test with the 1D solver ($\times$ symbols) and the 3D  line solver.
This figure shows cuts along the $x$, $y$, and $z$ axes of the
3D grid for a grid with $n_x=n_y=n_z=2*64+1$ spatial points.
The ordinate axis shows the coordinates, the 
$y$ axis the $\log$ of the  mean intensity averaged over the line 
profiles ($\Jb$) for cuts along the axes of the 3D grid. For the 1D 
comparison case the ordinate shows $\pm$ distance from the center.}
\end{figure}

In this test we have set $\epsilon_l=1$ to test the accuracy of the formal
solution by comparing to the results of the 1D code.  The 1D solver uses 64
radial points, distributed logarithmically in optical depth.  In
Fig.~\ref{fig:LTE:distance} we show the line mean intensities $\Jb$ as function
of distance from the center for both the 1D ($+$ symbols) and the 3D solver.
The results plotted in Fig.~\ref{fig:LTE:axes} show an excellent agreement
between the two solutions, showing that the line 3D RT formal solution is
comparable in accuracy with the corresponding 1D formal solution.  Note that
the difference of the distribution of spatial points (linear in the 3D case,
approximately logarithmic in the 1D case) causes a much lower resolution of the
3D calculations in the central regions and in a higher resolution of the 3D
calculations  in the outer regions compared to the 1D test case.

\subsection{Tests with line scattering}
\begin{figure}
\centering
\resizebox{\hsize}{!}{\includegraphics[angle=90]{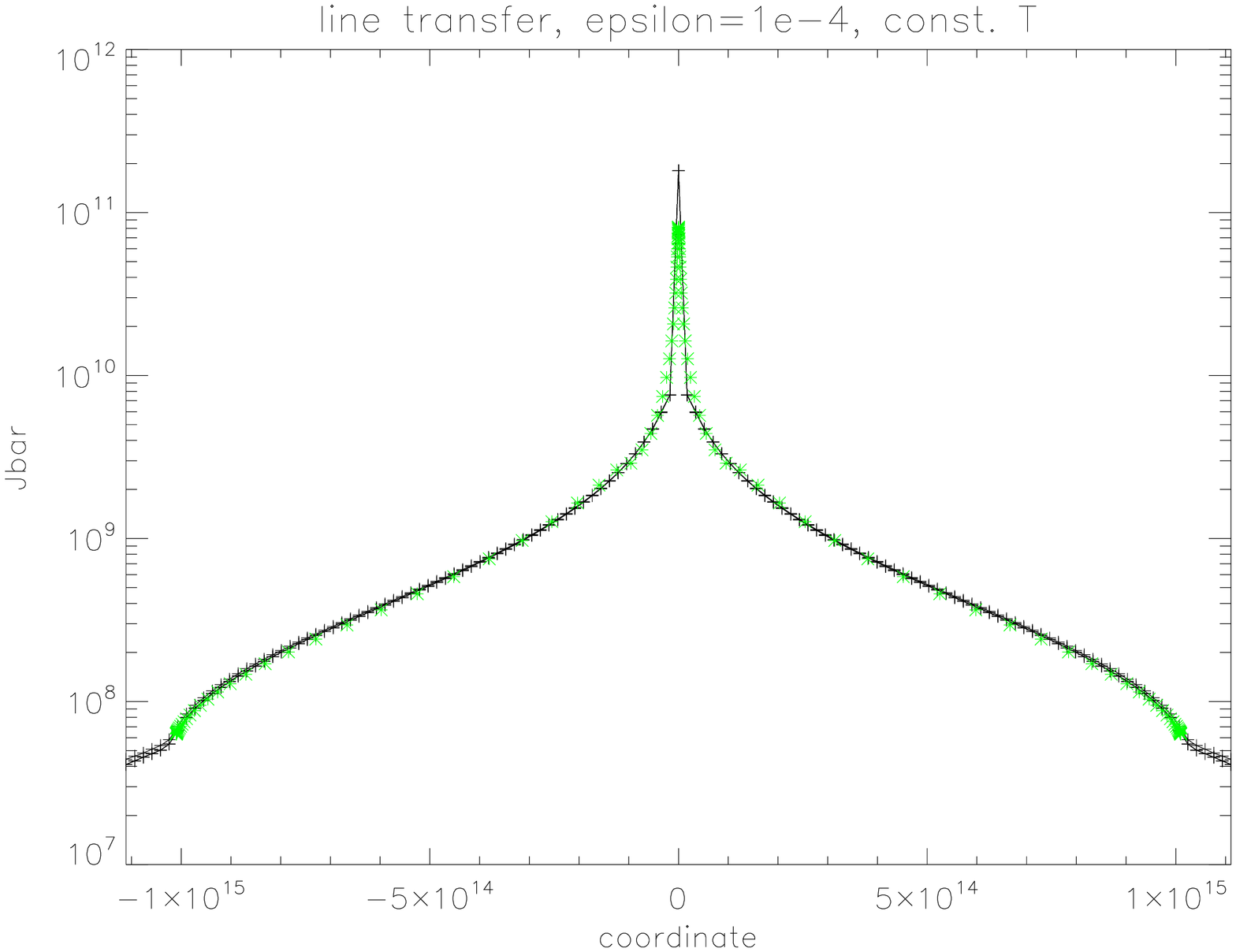}}
\caption{\label{fig:eps=-4:flat:axes} Comparison of the results obtained for the
$\epsilon_l=10^{-4}$ line  test (constant $T$)  with the 1D solver ($\times$ symbols) and the 3D
line solver.  This figure shows cuts along the $x$, $y$, and $z$ axes of the 3D
grid with $n_x=n_y=n_z=2*64+1$ spatial points.
The ordinate axis shows the coordinates, the 
$y$ axis the $\log$ of the  mean intensity averaged over the line 
profiles ($\Jb$) for cuts along the axes of the 3D grid. For the 1D 
comparison case the ordinate shows $\pm$ distance from the center.}
\end{figure}

\begin{figure}
\centering
\resizebox{\hsize}{!}{\includegraphics[angle=90]{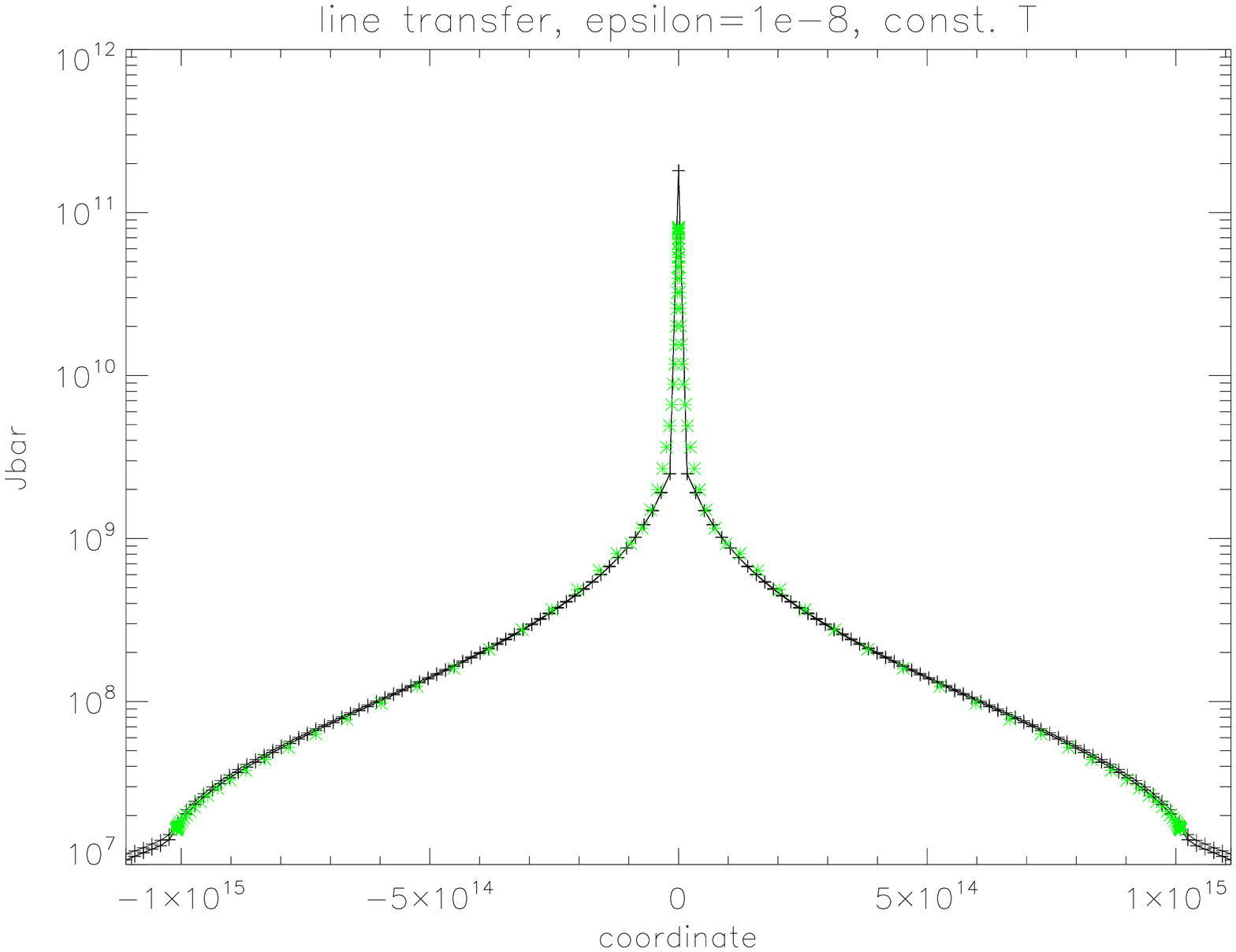}}
\caption{\label{fig:eps=-8:flat:axes} Comparison of the results obtained for the
$\epsilon_l=10^{-8}$ line  test (constant $T$)  with the 1D solver ($\times$ symbols) and the 3D
line solver.  This figure shows cuts along the $x$, $y$, and $z$ axes of the 3D
grid with $n_x=n_y=n_z=2*64+1$ spatial points.
The ordinate axis shows the coordinates, the 
$y$ axis the $\log$ of the  mean intensity averaged over the line 
profiles ($\Jb$) for cuts along the axes of the 3D grid. For the 1D 
comparison case the ordinate shows $\pm$ distance from the center.}
\end{figure}

We have run a number of test calculation similar to the LTE case but with line
scattering included. In Fig.~\ref{fig:eps=-4:flat:axes} we show the results for
$\epsilon_l=10^{-4}$ and in Fig.~\ref{fig:eps=-8:flat:axes} we show the results
for $\epsilon_l=10^{-8}$ as examples. In both cases, the dynamical ranges of
$\Jb$ are much larger than in the LTE case. The 3D calculations compare very
well to the 1D calculations, in particular in the outer zones. In the inner
parts the resolution of the 3D models is substantially lower than for the 1D
models, therefore the differences are largest there. In
Figs.~\ref{fig:eps=-4:flat:axes} and \ref{fig:eps=-8:flat:axes} we show the
results for a test model with a grey temperature structure. In these models,
the 3D spatial grid was substantially larger ($n_x=n_y=n_z=2*96+1$) in order to help
resolve the inner regions where the temperature gradients are very large. As
expected, the agreement is not as good in the inner regions as in the models
with a constant temperature, however, the models agree very well in the outer
regions. Overall, the agreement is very similar in quality compared to the case
with constant temperatures. {The differences in the mean intensity $\bar J$
between the 1D comparison case and the 3D case is several per-cent in the innermost
layers where the grid of the 3D case is under-sampled, the differences are
below 0.1\% in the outer zones.}

\subsection{Convergence}

\begin{figure*}
\centering
\includegraphics[width=12cm,angle=90]{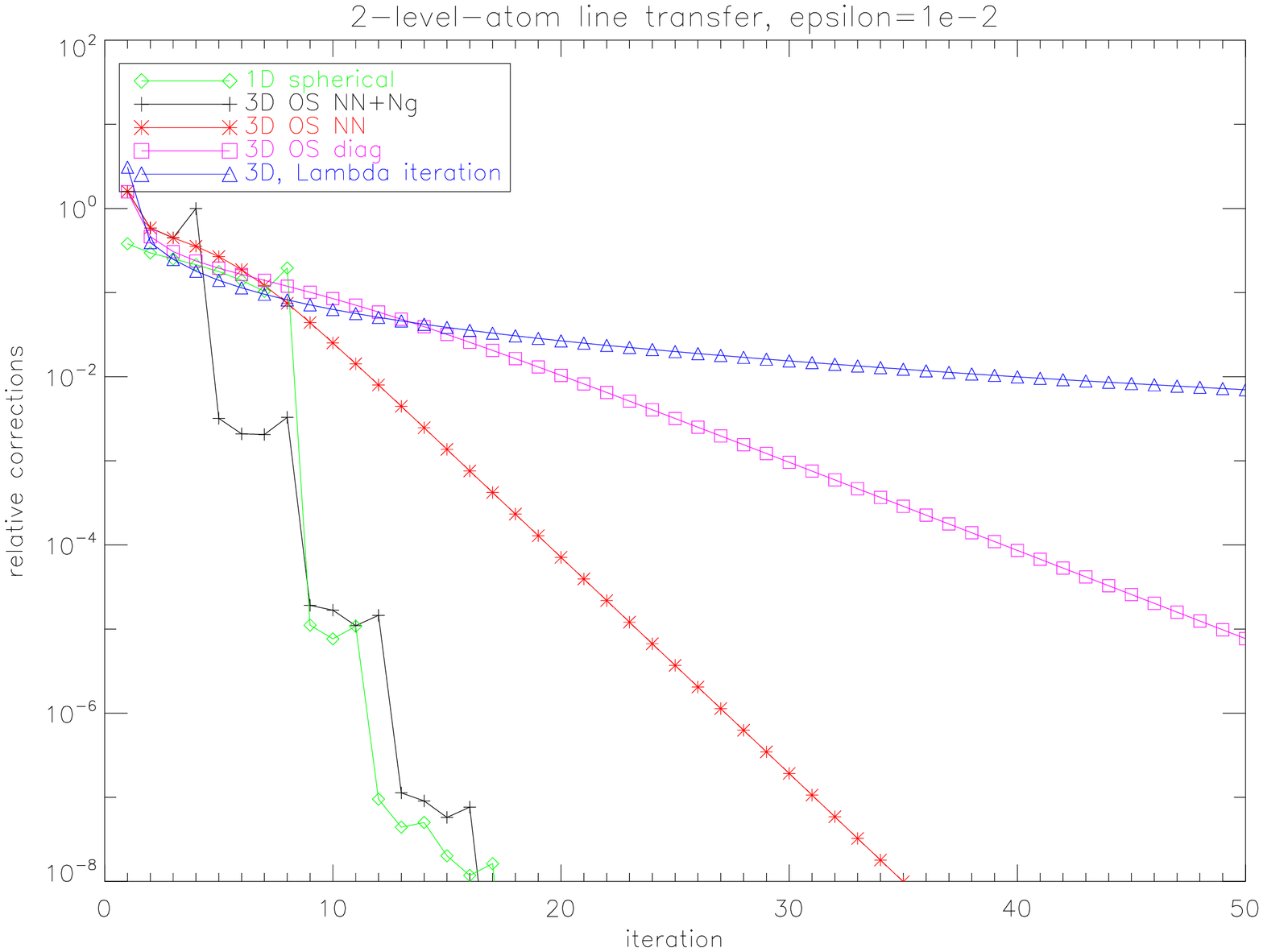}
\caption{\label{fig:convergence:eps2} Convergence of the iterations for
the line transfer case with $\epsilon_l=10^{-2}$. The maximum relative 
corrections (taken over all spatial points) are plotted vs.\ iteration
number.}
\end{figure*}

\begin{figure*}
\centering
\includegraphics[width=12cm,angle=90]{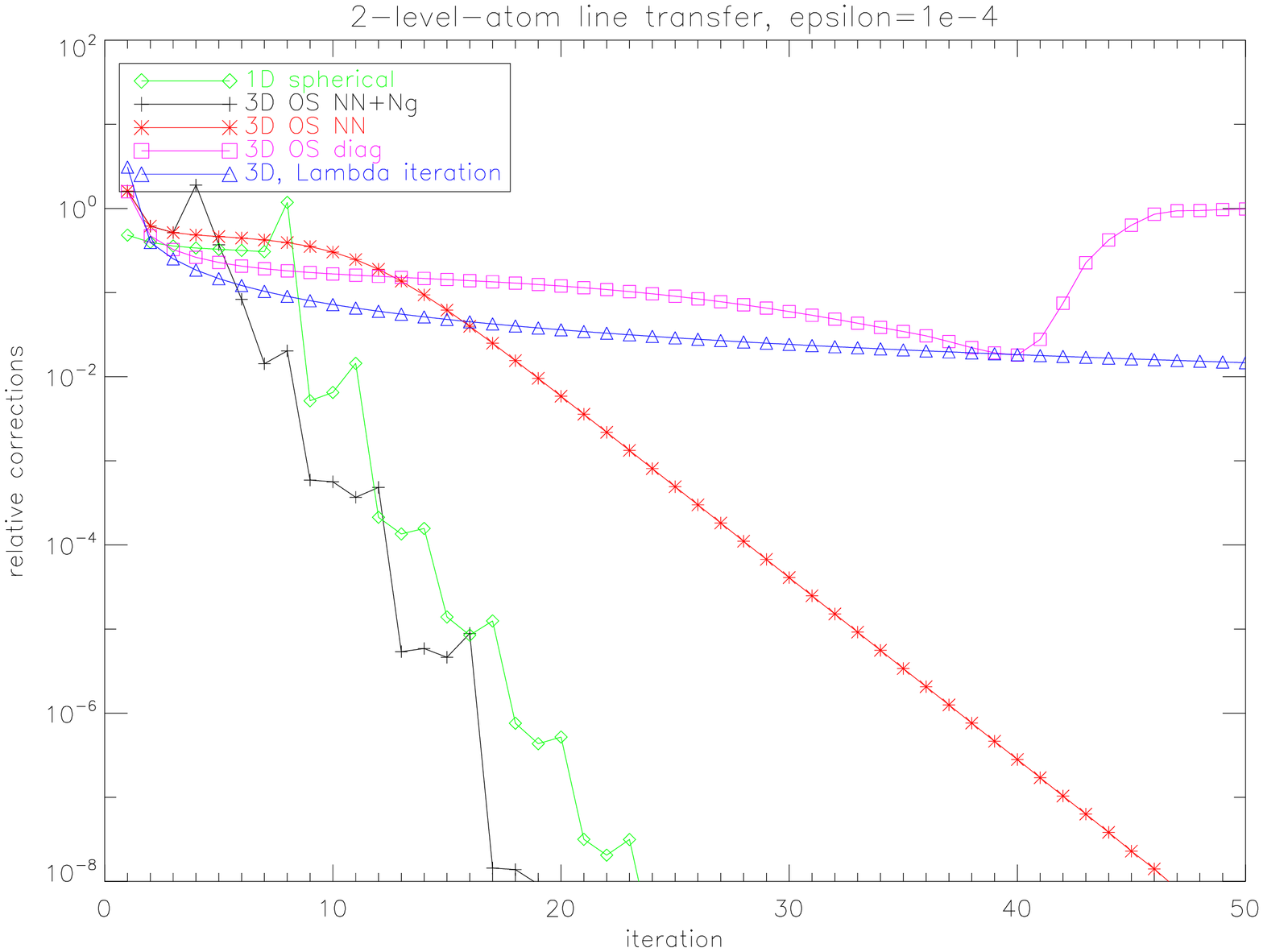}
\caption{\label{fig:convergence:eps4} Convergence of the iterations for
the line transfer case with $\epsilon_l=10^{-4}$. The maximum relative 
corrections (taken over all spatial points) are plotted vs.\ iteration
number.}
\end{figure*}

\begin{figure*}
\centering
\includegraphics[width=12cm,angle=90]{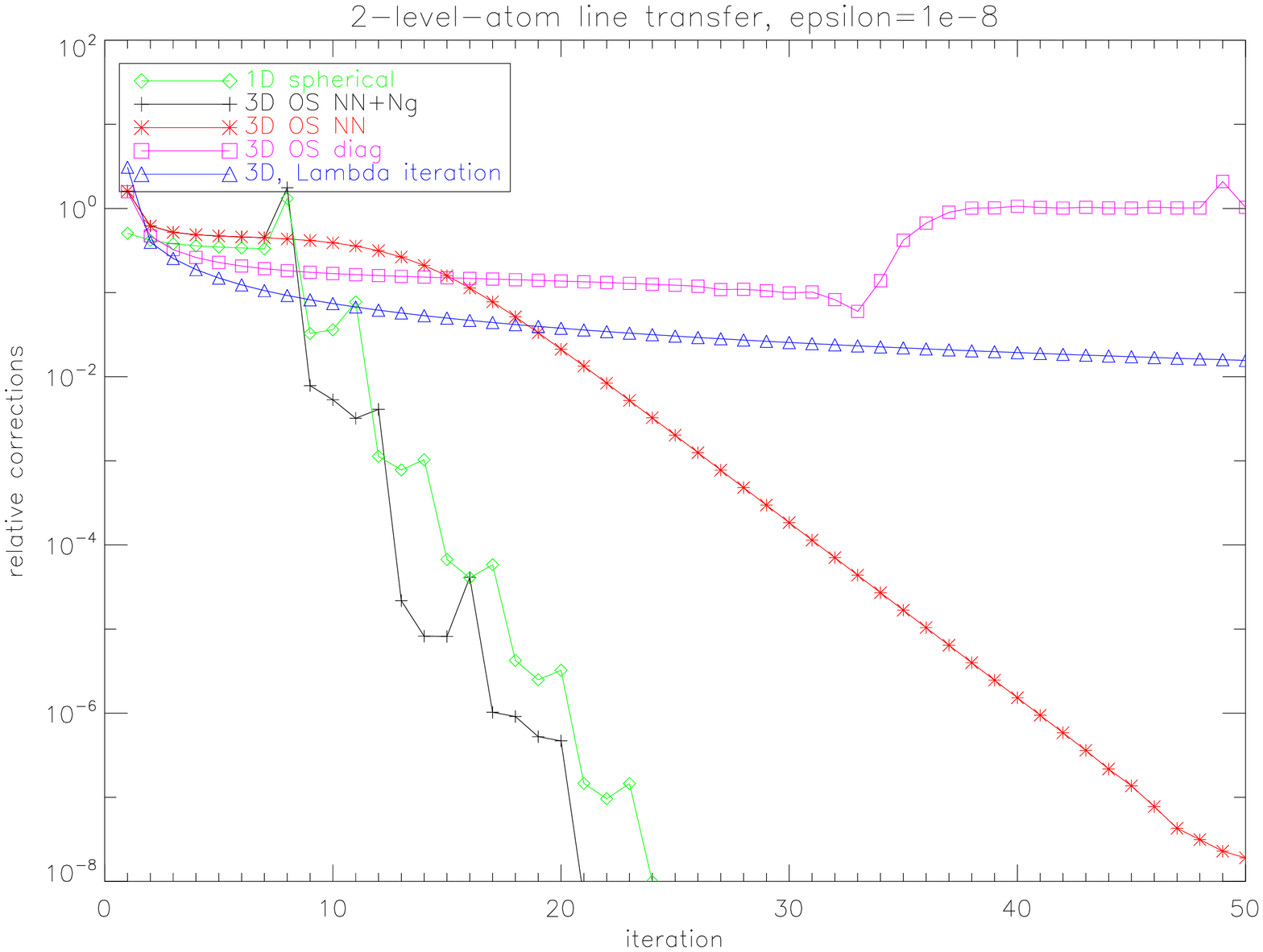}
\caption{\label{fig:convergence:eps8} Convergence of the iterations for
the line transfer case with $\epsilon_l=10^{-8}$. The maximum relative 
corrections (taken over all spatial points) are plotted vs.\ iteration
number.}
\end{figure*}

\begin{figure*}
\centering
\includegraphics[width=12cm,angle=90]{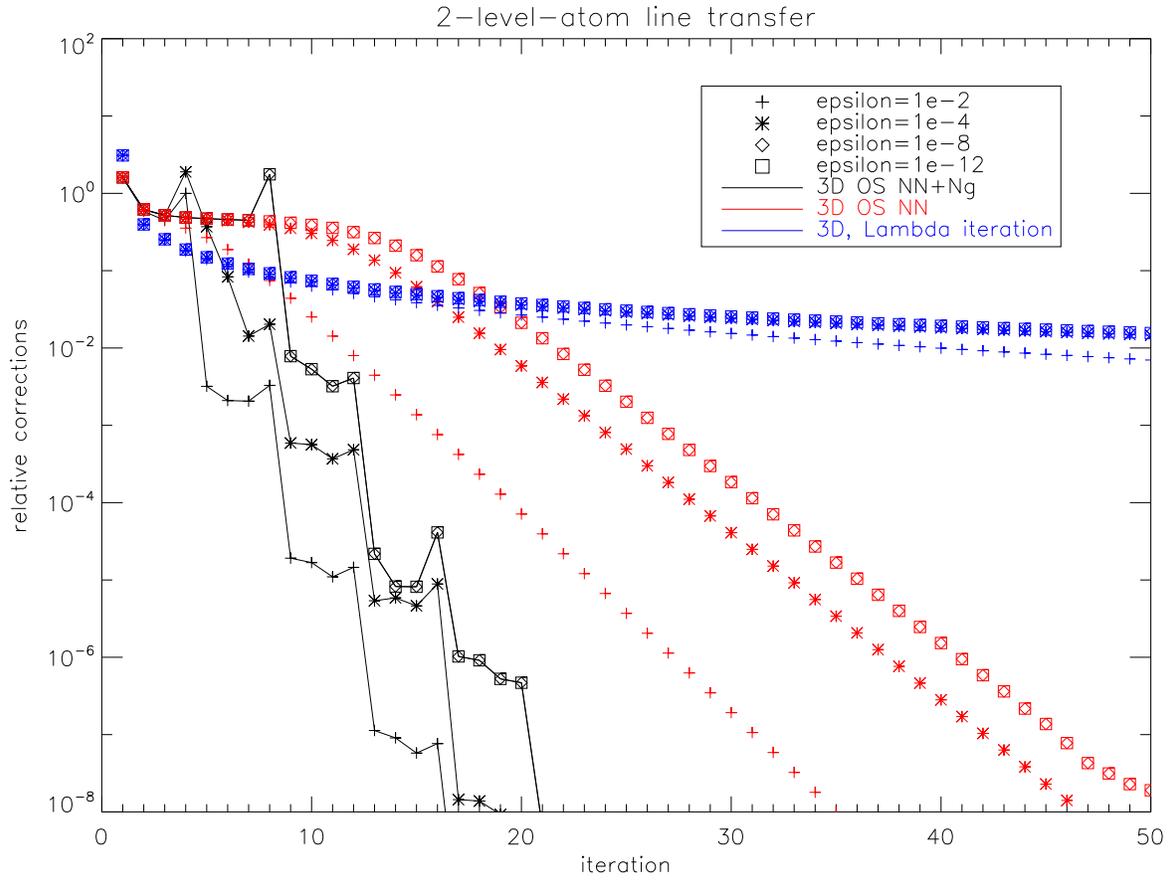}
\caption{\label{fig:convergence:eps_all} Convergence of the iterations for
the line transfer case with different $\epsilon_l$ as indicated in
the legend. The maximum relative 
corrections (taken over all spatial points) are plotted vs.\ iteration
number. The symbols without connecting lines are the convergence rates 
obtained without using Ng acceleration.}
\end{figure*}

{The convergence properties of the line transfer tests presented here are shown
in Figs.~\ref{fig:convergence:eps2}--\ref{fig:convergence:eps_all}. In each
figure, we show the convergence rates, as measured by the relative corrections
per iteration, for a number of test runs. In all tests show here we have used
$n_x=n_y=n_z=2*32+1$ points and $n_\theta=n_\phi=32$ solid angle points for the
3D test case and 64 radial points for the 1D comparison test. The iterations
were started with $S_l=\Jb=B$ at all spatial points, this initial guess causes
a relative error of about  $10$ in $\Jb$ at the outer zones for the case with
$\epsilon_l=10^{-2}$ and about $10^4$ in $\Jb$ at the outer zones for the case
with $\epsilon_l=10^{-8}$.  The plots show that the $\Lambda$ iteration is
useless even for the relatively benign case of $\epsilon_l=10^{-2}$.  The
operator splitting method delivers much larger corrections and is substantially
accelerated by the Ng method, similar to the results shown in Paper I. The
nearest-neighbor operator gives substantially better convergence rates than
the diagonal operator, cf.\ Fig~\ref{fig:convergence:eps2}, for the test
cases with with $\epsilon_l < 10^{-2}$ the convergence behavior of the diagonal
operator is unstable, the corrections tend to show oscillations. The nearest-neighbor
operator shows stable convergence with quickly declining corrections for 
all test cases, its convergence rate can be accelerated with Ng's method.
The total number of iterations required for the nearest-neighbor operator
is essentially identical to the 1D case with a tri-diagonal operator.}

\begin{table}
\caption{Scaling results for different parallel configurations. $N_{\rm worker}$
is the number of MPI processes working on a formal solution for a single
wavelength, $N_{\rm cluster}$ is the number of $N_{\rm worker}$ sets of 
processes working on different wavelength, the total number of MPI processes
is $N_{\rm cluster}\times N_{\rm worker}$. The column 'FS+$\Lstar$+OS step' gives
teh wallclock time (in seconds) for the calculation of the first formal solution plus
the construction of the $\Lstar$ operator plus the time for the first operator
splitting step, the column 'FS+OS step' is the time for the second (and subsequent)
formal solution and operator splitting step.}
\label{table:scaling}
\centering
\begin{tabular}{r r r r}
\hline\hline
$N_{\rm worker}$ & $N_{\rm cluster}$ & FS+$\Lstar$+OS step & FS+OS step \\
\hline
128 &  1 &	3018 & 	1143 \\
64  &  2 &	2595 &	1072 \\
32  &  4 &	2340 & 	1032 \\
16  &  8 &	2308 & 	1018 \\
 8  & 16 &	2264 & 	1052 \\
 4  & 32 &	2318 & 	1054 \\
\hline
\end{tabular}
\end{table}

\subsection{Parallelization}

We have implemented a hierarchical parallelization scheme on
distributed memory machines using the MPI framework similar to the
scheme discussed in \cite{parapap2}. Basically, the most efficient
parallelization opportunities in the problem are the solid angle and
wavelength sub-spaces. The total number of solid angles in the test
models is at least 4096, the number of wavelength points is 32 in the
tests presented here but will be much larger in large scale
applications. Thus even in the simple tests presented here, the
calculations could theoretically be run on 131072 processors. The work
required for each solid angle is roughly constant (the number of
points that need to be calculated depends on the angle points) and
during the formal solution process the solid angles are independent
from each other. The mean intensities (and other solid angle
integrated quantities) are only needed after the formal solution for
all solid angles is complete, so a single collective MPI operation is
needed to finish the computation of the mean intensities at each
wavelength.  Similarly, the wavelength integrated mean intensities
$\Jb$ are needed only after the formal solutions are completed for all
wavelengths (and solid angles). Therefore, the different wavelength
points can be computed in parallel also with the only communication
occurring as collective MPI operations after all wavelength points have
been computed. We thus divide the total number of processes up in a
number of `wavelength clusters' (each working on a different set of
wavelength points) that each have a number of `worker' processes which
work on a different set of solid angle points for any wavelength. In
the simplest case, each wavelength cluster has the same number of
worker processes so that $N_{\rm tot} = N_{\rm cluster}\times N_{\rm
  worker}$ where $N_{\rm tot}$ is the total number of MPI processes,
$N_{\rm cluster}$ is the number of wavelength clusters and $N_{\rm
  worker}$ is the number of worker processes for each wavelength
cluster. For our tests we could use a maximum number of 128 CPUs on
the HLRN IBM Regatta (Power4 CPUs) system. In
Table~\ref{table:scaling} we show the results for the 3 combinations
that we could run (due to computer time limitations) for a
$\epsilon_l=10^{-4}$ line transfer test case with 32 wavelength
points, $n_x=n_y=n_z=2*64+1$ spatial points and $n_\theta=n_\phi=64$
solid angle points. For example, the third row in the table is for a
configuration with $4$ wavelength clusters, each of them using $32$
CPUs working in parallel on different solid angles, for a total of 128
CPUs. The 3rd and 4th columns give the time (in seconds) for a full
formal solution, the construction of the $\lstar$ operator and an OS
step (the first iteration) and the time for a formal solution and an
OS step (the second iteration), respectively.  As the $\lstar$ has to
be constructed only in the first iteration, the overall time per
iteration drops in subsequent iterations.  Similarly to the 1D case,
the construction of the $\lstar$ is roughly equivalent to one formal
solution. We have verified that all parallel configurations lead to
identical results. The table shows that configurations with more
clusters are slightly more efficient, mostly due to better load
balancing.  However, the differences are not really significant in
practical applications so that the exact choice of the setup is not
important.  This also means that the code can easily scale to much
larger numbers of processors since realistic applications will require
much more than the 32 wavelength points used in the test
calculations. Note that the MPI parallelization can be combined with
shared memory parallelization (e.g., using openMP) in order to more
efficiently utilize modern multi-core processors with shared
caches. Although this is implemented in the current version of the 3D
code, we do not have access to a machine with such an architecture and
it was not efficient to use openMP on multiple single core processors.

\section{Conclusions}

Using rather difficult test problems, we have shown that our 3D
long-characteristics method gives very good results when compared to our well-tested
1D code. The main differences are due to poorer spatial resolution close to the
center of the grid in the 3D case.  The code has also been parallelized and
scales to very large numbers of processors. Future work will examine a
short-characteristics method of solution which should enable higher resolution
grids with the same memory requirements and an extension to full multi-level 
NLTE modeling.  

\begin{acknowledgements}
  This work was supported in part by by NASA grants NAG5-12127 and
  NNG04GD368, and NSF grants AST-0204771 and AST-0307323.  Some of the
  calculations presented here were performed at the H\"ochstleistungs
  Rechenzentrum Nord (HLRN); at the NASA's Advanced Supercomputing
  Division's Project Columbia, at the Hamburger Sternwarte Apple G5
  and Delta Opteron clusters financially supported by the DFG and the
  State of Hamburg; and at the National Energy Research Supercomputer
  Center (NERSC), which is supported by the Office of Science of the
  U.S.  Department of Energy under Contract No. DE-AC03-76SF00098.  We
  thank all these institutions for a generous allocation of computer
  time.
\end{acknowledgements}

\bibliography{yeti,radtran,rte_paper2}
\clearpage

\begin{figure}
\centering
\caption{\label{fig:LTE:distance} Comparison of the results obtained 
for the LTE line  test with the 1D solver ($+$ symbols) and the 3D  line solver.
The $x$ axis shows the distances from the center of the sphere, the 
$y$ axis the $\log$ of the  mean intensity averaged over the line 
profiles ($\Jb$). The 3D model uses $n_x=n_y=n_z=2*64+1$ spatial points.}
\end{figure}

\begin{figure}
\centering
\resizebox{\hsize}{!}{\includegraphics[angle=90]{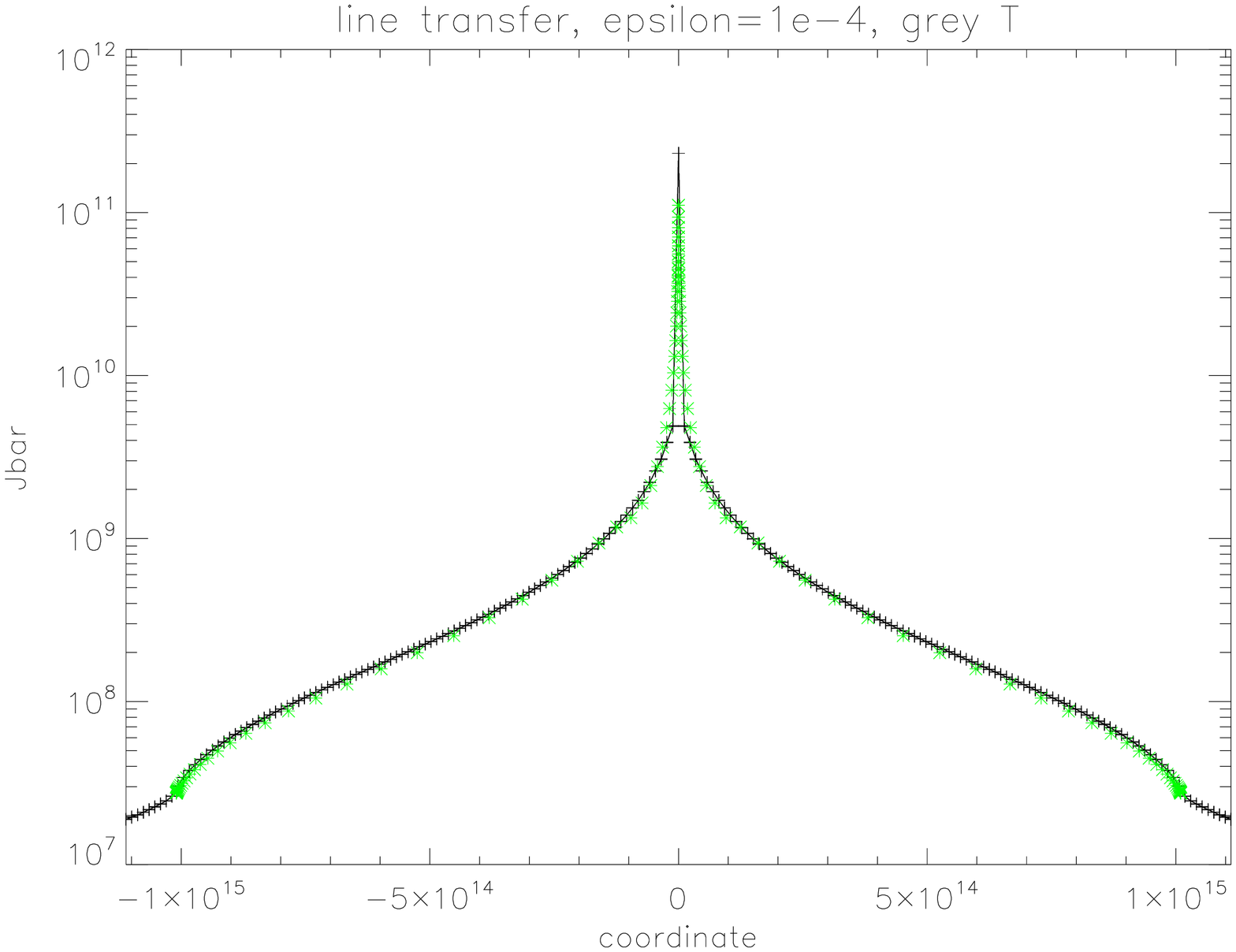}}
\caption{\label{fig:eps=-4:axes} Comparison of the results obtained for the
$\epsilon_l=10^{-4}$ line  test (grey $T$)  with the 1D solver ($\times$ symbols) and the 3D
line solver.  This figure shows cuts along the $x$, $y$, and $z$ axes of the 3D
grid with $n_x=n_y=n_z=2*96+1$ spatial points.
The ordinate axis shows the coordinates, the 
$y$ axis the $\log$ of the  mean intensity averaged over the line 
profiles ($\Jb$) for cuts along the axes of the 3D grid. For the 1D 
comparison case the ordinate shows $\pm$ distance from the center.}
\end{figure}

\begin{figure}
\centering
\resizebox{\hsize}{!}{\includegraphics{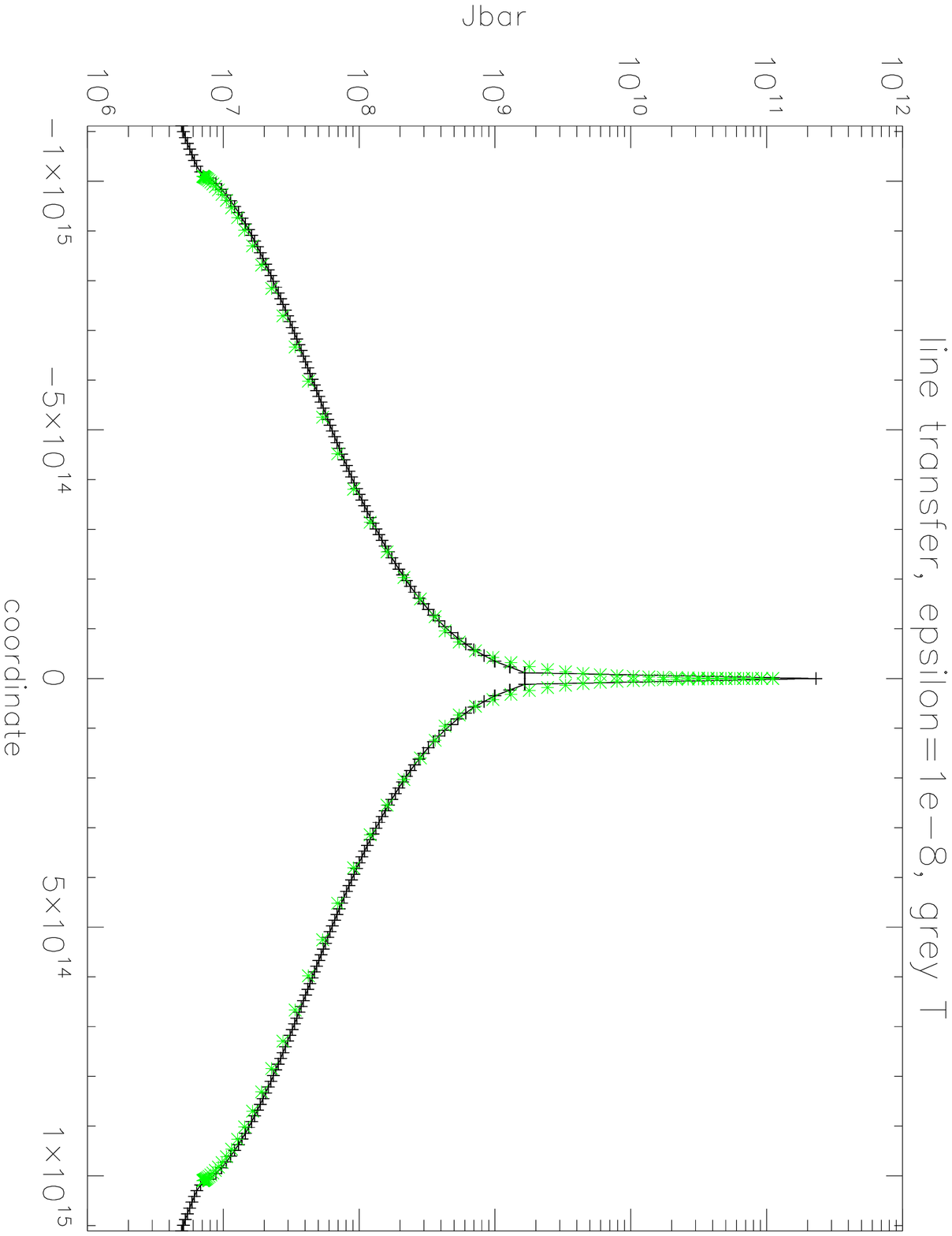}}
\caption{\label{fig:eps=-8:axes} Comparison of the results obtained for the
$\epsilon_l=10^{-8}$ line  test (grey $T$)  with the 1D solver ($\times$ symbols) and the 3D
line solver.  This figure shows cuts along the $x$, $y$, and $z$ axes of the 3D
grid  with $n_x=n_y=n_z=2*64+1$ spatial points.
The ordinate axis shows the coordinates, the 
$y$ axis the $\log$ of the  mean intensity averaged over the line 
profiles ($\Jb$) for cuts along the axes of the 3D grid. For the 1D 
comparison case the ordinate shows $\pm$ distance from the center.}
\end{figure}

\end{document}

%% file: macros.tex
%
%
\def\valid{}    

\font\caps=cmcsc10                  
\font\dunh=cmdunh10  at 12.0 true pt 
\font\dunhs=cmdunh10 
\font\vbold=cmbx10 scaled \magstep1 
\font\sevenbf=cmbx7
\font\sevenit=cmti7
\font\Kapi=cmr17

\def\MEV{DOME}
\def\RTE{equation of radiative transfer}
\def\etal{{et al}}
\def\HW{H\&W}
\def\OK{O\&K}
\def\ok{O\&K}
\def\RH{R\&H}

\def\ibmrs{\hbox{\tt RS/6000}}
\def\hp{\hbox{\tt HP~9000}}
\def\dec{\hbox{\tt DEC~5000}}
\def\axp{\hbox{\tt AXP}}
\def\ibmmf{\hbox{\tt IBM~3090}}
\def\ibmpc{\hbox{\tt 486DX}}
\def\cray{\hbox{\tt Cray 2}}
\def\ymp{\hbox{\tt YMP}}
\def\nec{\hbox{\tt NEC}}

\def\g{\gamma}
\def\b{\beta}
\def\m{\mu}
\def\e{\epsilon}
\def\n{\nu}
\def\l{\lambda}
\def\L{\Lambda}
\def\t{\tau}
\def\pder#1#2{{\partial #1 \over \partial #2}}
\def\div#1#2{{#1\over #2}}
\def\rout{\ifmmode{r_{\rm out}}\else\hbox{$r_{\rm out}$}\fi}
\def\tmax{\ifmmode{\tau_{\rm max}}\else\hbox{$\tau_{\rm max}$}\fi}
\def\tstd{\ifmmode{\tau_{\rm std}}\else\hbox{$\tau_{\rm std}$}\fi}
\def\vmax{\ifmmode{v_{\rm max}}\else\hbox{$v_{\rm max}$}\fi}
\def\muE{\ifmmode{\mu_{\rm E}}\else\hbox{$\mu_{\rm E}$}\fi} 
\def\pE{\ifmmode{p_{\rm E}}\else\hbox{$p_{\rm E}$}\fi} 
\def\bmax{\ifmmode{\b_{\rm max}}\else\hbox{$\b_{\rm max}$}\fi}
\def\kms{\hbox{$\,$km$\,$s$^{-1}$}}
\def\ergs{\hbox{$\,$erg$\,$s$^{-1}$}}
\def\kpc{\hbox{$\,$kpc} }
\def\ang{\hbox{\AA}}
\def\Msun{\hbox{$\,$M$_\odot$} }
\def\Lsun{\hbox{$\,$L$_\odot$} }
\def\Teff{\hbox{$\,T_{\rm eff}$} }
\def\alog#1{\times 10^{#1}}
\def\rin{\hbox{$r_{\rm in}$} }
\def\rout{\hbox{$r_{\rm out}$} }

\def\lstar{\ifmmode{\Lambda^*}\else\hbox{$\Lambda^*$}\fi} 
\def\Lstar{\ifmmode{\Lambda^*}\else\hbox{$\Lambda^*$}\fi} 
\def\Rop{\ifmmode{[R_{ij}]}\else\hbox{$[R_{ij}]$}\fi}
\def\Rij{\Rop}
\def\Rji{\ifmmode{[R_{ji}]}\else\hbox{$[R_{ji}]$}\fi}
\def\Rstar{\ifmmode{[R_{ij}^*]}\else\hbox{$[R_{ij}^*]$}\fi}
\def\Rijstar{\Rstar}
\def\Rjistar{\ifmmode{[R_{ji}^*]}\else\hbox{$[R_{ji}^*]$}\fi}
\def\DRji{\ifmmode{[\Delta R_{ji}]}\else\hbox{$[\Delta R_{ji}]$}\fi}
\def\DRij{\ifmmode{[\Delta R_{ij}]}\else\hbox{$[\Delta R_{ij}]$}\fi}

\def\Jb{{\bar J}}
\def\Jnew{{\bar J_{\rm new}}}
\def\Jold{{\bar J_{\rm old}}}
\def\Jfs{{\bar J_{\rm fs}}}
\def\Snew{{S_{\rm new}}}
\def\Sold{{S_{\rm old}}}
\def\Amat{\mat{A}}             

\def\ns{\ifmmode{N_{\rm s}}          
        \else\hbox{$N_{\rm s}$}\fi}
\def\ion#1{\hbox{ #1}}         

\def\peq{\mathbin{\hbox{$+$}\hbox{$=$}}}

\def\mat#1{{\bf #1}}     
\def\vek#1{{#1}}         

\newcount\eqcount
\eqcount=0
\def
  \nummer{
    \global\advance\eqcount by 1
    (\the\eqcount)
  }

\def
  \numadv{
    \global\advance\eqcount by 1
  }

\def
   \numout#1{
     (\the\eqcount #1)
  }

\def\ivek#1#2{\ifmmode{\vek{I}^{#1}_{#2}}
        \else\hbox{$\vek{I}^{#1}_{#2}$}\fi}

\def\ip#1{\ivek{+}{#1}}      
\def\im#1{\ivek{-}{#1}}      

\def\tmat#1#2{\ifmmode{\mat{t}^{#1}_{#2}}
        \else\hbox{$\mat{t}^{#1}_{#2}$}\fi}
\def\rmat#1#2{\ifmmode{\mat{r}^{#1}_{#2}}
        \else\hbox{$\mat{r}^{#1}_{#2}$}\fi}
\def\bvek#1#2{\ifmmode{\beta^{#1}_{#2}}
        \else\hbox{$\beta^{#1}_{#2}$}\fi}

\def\tpi#1{\tmat{+}{#1}}
\def\tmi#1{\tmat{-}{#1}}
\def\rmi#1{\rmat{-}{#1}}
\def\rpi#1{\rmat{+}{#1}}
\def\bpi#1{\bvek{+}{#1}}
\def\bmi#1{\bvek{-}{#1}}

\def\tp{\tmat{+}{}}          
\def\tm{\tmat{-}{}}          
\def\rmm{\rmat{-}{}}         
\def\rp{\rmat{+}{}}          
\def\bp{\bvek{+}{}}          
\def\bm{\bvek{-}{}}          
\def\tpm{\tmat{\pm}{}}       
\def\rpm{\rmat{\pm}{}}       
\def\bpm{\bvek{\pm}{}}       

\def\lp{\ifmmode{\lambda^+_\tau}           
        \else\hbox{$\lambda^+_\tau$}\fi}
\def\lm{\ifmmode\lambda^-_\tau             
        \else\hbox{$\lambda^-_\tau$}\fi}

%% file: aas_journals.tex
%
%
%
%



\def\aasref@jnl#1{{\rm #1}}

\def\aj{\aasref@jnl{AJ}}                   
\def\araa{\aasref@jnl{ARA\&A}}             
\def\apj{\aasref@jnl{ApJ}}                 
\def\apjl{\aasref@jnl{ApJ}}                
\def\apjs{\aasref@jnl{ApJS}}               
\def\ao{\aasref@jnl{Appl.~Opt.}}           
\def\apss{\aasref@jnl{Ap\&SS}}             
\def\aap{\aasref@jnl{A\&A}}                
\def\aapr{\aasref@jnl{A\&A~Rev.}}          
\def\aaps{\aasref@jnl{A\&AS}}              
\def\azh{\aasref@jnl{AZh}}                 
\def\baas{\aasref@jnl{BAAS}}               
\def\jrasc{\aasref@jnl{JRASC}}             
\def\memras{\aasref@jnl{MmRAS}}            
\def\mnras{\aasref@jnl{MNRAS}}             
\def\pra{\aasref@jnl{Phys.~Rev.~A}}        
\def\prb{\aasref@jnl{Phys.~Rev.~B}}        
\def\prc{\aasref@jnl{Phys.~Rev.~C}}        
\def\prd{\aasref@jnl{Phys.~Rev.~D}}        
\def\pre{\aasref@jnl{Phys.~Rev.~E}}        
\def\prl{\aasref@jnl{Phys.~Rev.~Lett.}}    
\def\pasp{\aasref@jnl{PASP}}               
\def\pasj{\aasref@jnl{PASJ}}               
\def\qjras{\aasref@jnl{QJRAS}}             
\def\skytel{\aasref@jnl{S\&T}}             
\def\solphys{\aasref@jnl{Sol.~Phys.}}      
\def\sovast{\aasref@jnl{Soviet~Ast.}}      
\def\ssr{\aasref@jnl{Space~Sci.~Rev.}}     
\def\zap{\aasref@jnl{ZAp}}                 
\def\nat{\aasref@jnl{Nature}}              
\def\iaucirc{\aasref@jnl{IAU~Circ.}}       
\def\aplett{\aasref@jnl{Astrophys.~Lett.}} 
\def\apspr{\aasref@jnl{Astrophys.~Space~Phys.~Res.}}
\def\bain{\aasref@jnl{Bull.~Astron.~Inst.~Netherlands}} 
\def\fcp{\aasref@jnl{Fund.~Cosmic~Phys.}}  
\def\gca{\aasref@jnl{Geochim.~Cosmochim.~Acta}}   
\def\grl{\aasref@jnl{Geophys.~Res.~Lett.}} 
\def\jcp{\aasref@jnl{J.~Chem.~Phys.}}      
\def\jgr{\aasref@jnl{J.~Geophys.~Res.}}    
\def\jqsrt{\aasref@jnl{J.~Quant.~Spec.~Radiat.~Transf.}}
\def\memsai{\aasref@jnl{Mem.~Soc.~Astron.~Italiana}}
\def\nphysa{\aasref@jnl{Nucl.~Phys.~A}}   
\def\physrep{\aasref@jnl{Phys.~Rep.}}   
\def\physscr{\aasref@jnl{Phys.~Scr}}   
\def\planss{\aasref@jnl{Planet.~Space~Sci.}}   
\def\procspie{\aasref@jnl{Proc.~SPIE}}   

\let\astap=\aap
\let\apjlett=\apjl
\let\apjsupp=\apjs
\let\applopt=\ao